\documentclass[journal]{IEEEtran}
\usepackage[pdftex]{graphicx}
\usepackage[center]{caption}
\usepackage{setspace,flushend}
\usepackage{amssymb,amsmath,amsthm}
\usepackage[caption=false,font=footnotesize]{subfig}

\newcommand{\bC}{\mathbb C}
\newcommand{\bF}{\mathbb F}
\newcommand{\bK}{\mathbb K}
\newcommand{\bR}{\mathbb R}
\newcommand{\cA}{\mathcal A}
\newcommand{\cS}{\mathcal S}
\newcommand{\vc}{\mathbf c}
\newcommand{\vn}{\mathbf n}

\newcommand{\ve}{\mathbf e}
\newcommand{\vf}{\mathbf f}
\newcommand{\vx}{\mathbf x}
\newcommand{\vy}{\mathbf y}
\newcommand{\vz}{\mathbf z}
\newcommand{\mA}{\mathbf A}
\newcommand{\mC}{\mathbf C}

\newcommand{\mG}{\mathbf G}

\newtheorem{thm}{Theorem}[section]

\newtheorem{lem}[thm]{Lemma}
\newtheorem{cor}[thm]{Corollary}
\newtheorem{defn}{Definition}[section]

\begin{document}

\title{On Finite Alphabet Compressive Sensing}
\author{Abhik~Kumar~Das and Sriram~Vishwanath\\
Dept. of E.C.E., The University of Texas at Austin, U.S.A.\\
e-mail: akdas@utexas.edu, sriram@austin.utexas.edu}

\maketitle

\begin{abstract}
This paper considers the problem of compressive sensing over a finite alphabet, where the finite alphabet may be inherent to the nature of the data or a result of quantization. There are multiple examples of finite alphabet based static as well as time-series data with inherent sparse structure; and quantizing real values is an essential step while handling real data in practice. We show that there are significant benefits to analyzing the problem while incorporating its finite alphabet nature, versus ignoring it and employing a conventional real alphabet based toolbox. Specifically, when the alphabet is finite, our techniques (a) have a lower sample complexity compared to real-valued compressive sensing for sparsity levels below a threshold; (b) facilitate constructive designs of sensing matrices based on coding-theoretic techniques; (c) enable one to solve the exact $\ell_0$-minimization problem in polynomial time rather than a approach of convex relaxation followed by sufficient conditions for when the relaxation matches the original problem; and finally, (d) allow for smaller amount of data storage (in bits).
\end{abstract}


\section{Introduction}
Compressive sensing has witnessed an explosion of research and literature in recent years. It has found useful applications in diverse fields, ranging from signal processing and micro-arrays to imaging and computer vision \cite{Baran06,ParVikMisHas08,Romb08,DuaDavTakLasSunKelBar08}. The theory behind compressive sensing permits the sensing and recovery of signals, that are ``sparse" in some domain, using a small number of linear measurements, roughly proportional to the number of non-zero values the signals take in the sparse domain \cite{CandTao05,CandTao06,Dono06}. To be precise, a real-valued $n$-dimensional signal, with a $b$-sparse representation in some basis, can be captured using $m=O(b\log (n/b))$ measurements based on linear combinations of the signal values ($b,n,m\in\mathbb{N}$, $b<n$). As such, compressive sensing finds its utility in setups where there is an inherent sparse structure in the nature of data, and storing or collecting measurements can be expensive.

There are multiple practical algorithms for near-perfect recovery of real-valued sparse signals from their linear measurements, in the presence or in the absence of noise \cite{Tibs96,ChenDonoSaun98,CandesTao05,SarvBarBara06,DonoTsai06,TroppGilb07}. These algorithms tend to either be based on linear programming (like basis pursuit and Lasso) or low complexity iterative techniques (like orthogonal matching pursuit). Regardless of the algorithmic framework, a common underlying feature of real-valued compressive sensing is a property of incoherence in the sensing matrix corresponding to linear measurements (e.g., RIP), that serves as a sufficient condition for accurate reconstruction of sparse signals \cite{CandRom09,Wain09}. From an analytical perspective, there is a large and growing body of literature on the necessary and sufficient conditions for accurate recovery of sparse signals.

In practice, signals are not always real-valued. For example, opinion polls, ranking information, commodity sales numbers, and other counting data sets including arrivals at a queue/server are inherently discrete-valued. Moreover, some of what might otherwise be regarded as continuous-valued data sets are conventionally ``binned" into finite alphabet sets. This includes rainfall data, power generation data and many other examples where quantized values are of interest as the output of the sensing process. In such cases, knowledge of the nature of the alphabet can prove to be useful, which together with the underlying sparsity property can lead to alternate and efficient algorithms for finite alphabet compressive sensing.

In this paper, we consider a setup where the sensed information belongs to a known finite alphabet. We treat this alphabet as a subset of a suitable finite field, and make use of the field structural properties for compressing and reconstructing sparse signals. This approach enables us to use tools from algebraic coding theory to construct sensing matrices and design efficient algorithms for recovering sparse signals. In this process, we also build a deeper connection between the areas of algebraic coding theory and compressive sensing than what is currently understood in literature \cite{DrapMalek09,ZhanPfis08}.

Our main application domain in this paper is in tracking discrete-valued time-series data. For example, consider the time-series data corresponding to the backlog in a set of queues in a discrete-time system. This backlog is typically discrete-valued, and we assume the change in backlog from one time instance to the next has a sparse nature. In many queuing applications, it is practically infeasible to exactly measure and store the backlog in each queue due to extremely short timescales; therefore, an efficient compression mechanism for sensing and storage is desirable. Our goal is to track this time-series accurately. Our benchmark for comparison is real-valued compressive sensing, i.e., storing real-valued linear combination based measurements and using a convex relaxation approach for recovering the sparse differences between successive time instances. In Section \ref{sec:simulation}, we compare this approach to the one presented in this paper, and find that for the same number of samples, the real-valued approach suffers from error accumulation as the time-series progresses, while the finite alphabet approach tracks the values exactly.

\subsection{Motivation}
In case of real-valued compressive sensing, the recovery of sparse signals from linear measurements reduces to solving a $\ell_0$-minimization problem. To be precise, given $b$-sparse vector $\vx\in\mathbb{R}^{n\times 1}$, sensing matrix $\mA\in\mathbb{R}^{m\times n}$ and vector o measurements $\vy=\mA\vx$, $\vx$ can be recovered from $\vy$ by solving
\begin{equation}\label{eqn:sys}
\min||\vx||_0 \quad \textrm{s.t.} \quad\vy=\mA\vx.
\end{equation}
However this optimization problem is non-convex and known to be NP-hard for known meaningful constructions of sensing matrices. Therefore, a standard relaxation approach used in compressive sensing is to replace $\ell_0$-norm by $\ell_1$-norm, i.e., solve the convex optimization problem of $\ell_1$-minimization. A sufficient condition for exact recovery via $\ell_1$-minimization is that $\mA$ satisfies incoherence properties such as RIP.

Typically, finite alphabet problems are  more difficult or complicated to solve than their real-valued counterparts. For example, finding the logarithm is a simple inversion for real values but NP-hard over finite fields. This difficulty forms the basis of the well-known Diffie-Hellman key exchange algorithm, as do other hardness guarantees over finite fields. In our case, we have an interesting inversion of facts. \eqref{eqn:sys} is NP-hard for most sensing matrices of interest in compressive sensing over reals. However, using algebraic coding theory, we can design sensing matrices $\mA$ that are `good' for sensing as well as enable \eqref{eqn:sys} to be solved in polynomial-time.

Another important reason for finite alphabet analysis of compressive sensing is storage space. Real-values are analytical abstract artifacts, and are a blessing for mathematicians and applied mathematicians alike, but in practice, values must be stored and processed in form of discrete alphabet sets.  The requirement for storage space (in bits or any other unit) is an important area of concern for any compression framework, and applications of compressive sensing are no exception. We show that our methodology not only affords a lower sample complexity (in symbols) under certain settings, but it also has a lower storage requirement in terms of bits of information to be stored for exact reconstruction of sparse signals. Finally, although relaxations of optimization problems are immensely useful and mathematically elegant, being able to solve the original problem in polynomial time is always desirable. This, for example, enables accurate tracking of sparse-structured discrete time-series data using finite alphabet compressive sensing, while avoiding error accumulation with time.

%There exist multiple applications where low-resolution ADC's are used and/or on-the-fly compression might be necessary. In such situations, the use of minimal storage space and fast processing makes sense. The compression and recovery methods in the compressive sensing approach involve operations over reals and as such the amount of bits required for storage of symbols is the same as that required for reals. However, in the algebraic approach, the sensing matrix and source sparse vector can have the same resolution -- this gives potential storage space benefits if this resolution is less compared to the resolution used for storing real numbers in the setup.
\subsection{Related Work}
The fact that real-valued compressive sensing allows for recovery of sparse signals based on linear measurements is reminiscent of error correction in linear channel codes and compression by lossless source codes over finite alphabet or fields \cite{RyanLin09,Csis82}. Such similarities have been identified in existing literature to serve varied goals. For example, the use of bipartite expander graphs to design real-valued sensing matrices is investigated in \cite{XuHass07}. The connection between real-valued compressive sensing and linear channel codes is explored in \cite{ZhanPfis08}, by viewing sparse signal compression as syndrome-based source coding over real numbers and making use of linear codes over finite fields of large sizes. The design of real-valued sensing matrices based on LDPC codes is examined in \cite{SarvBarBara2006} and \cite{DimakVont09}. The connection between sparse learning problems and coding theory is studied in \cite{Mahdi11}. For real-valued compressive sensing over finite alphabet, the sparse signal recovery approaches that have been examined include approximate message passing \cite{Muller}, sphere decoding and semi-definite relaxation \cite{Tian}. However, an algebraic understanding of compressive sensing, particularly over finite fields, is yet limited, which is the main contribution of this paper.

\subsection{Main Results}\label{subsec:mainresults}
We use an algebraic framework for analyzing the recovery of sparse signals/vectors based on finite alphabet, given the set of linear measurements. We show that $m = \Theta( b \lceil\log_q n\rceil)$ measurements are sufficient for exact recovery of any $b$-sparse $n$-dimensional signal based on a finite alphabet of size $q$ -- this is smaller compared to the number of measurements needed for  real-valued compressive sensing for non-trivial ranges of $b,n,q$ (which ensures sparsity is below some threshold) \cite{CandRom09,Wain09}. We describe a coding-theoretic approach for constructing the sensing matrices. Note that this is a straightforward application of coding-theoretic principles, and thus the mathematical concepts are by no means new -- what makes it interesting is its connection and relevance to finite alphabet compressive sensing. We show the versatility and applicability of our approach to the case of noisy measurements, for both probabilistic and worst-case noise models. We apply our approach for tracking discrete-valued time-series with sparse changes, based on synthetic and real-world data; we find that, for the same number of samples, our approach performs accurate tracking while the real-valued approach accumulates error and drift in values as the time-series progresses.

We wish to emphasize that, even if the mathematical tools we use are known, our results are not obvious -- for example, there is no obvious reason why the bounds on sample complexity for the discrete case should be different (or better) than that of the real-valued case. Indeed, intuition suggests that discrete analysis should always be inferior in sample complexity, as discrete is a special case and more restrictive than its real-valued counterpart. Our contribution lies in the application of these known tools to finite alphabet data with sparse structure, and realizing that, for certain cases, lossless compression is possible with fewer samples and lesser storage space than its real-valued counterpart, and that it finds natural application in tracking discrete time-series data with sparse changes.

The rest of the paper is organized as follows. We describe the preliminaries in Section \ref{sec:prelims}, that introduces the problem setup and provides background on the relevant algebraic concepts used for analysis. We examine the problem of finite alphabet compressive sensing for the cases of noiseless and noisy measurements in Sections \ref{sec:noiseless} and \ref{sec:noisy} respectively. We demonstrate the simulation results based on our approach in Section \ref{sec:simulation}, and conclude the paper with Section \ref{sec:conclusion}.

\section{Preliminaries}\label{sec:prelims}
\begin{figure}[t]
\centering
\includegraphics[scale=0.37]{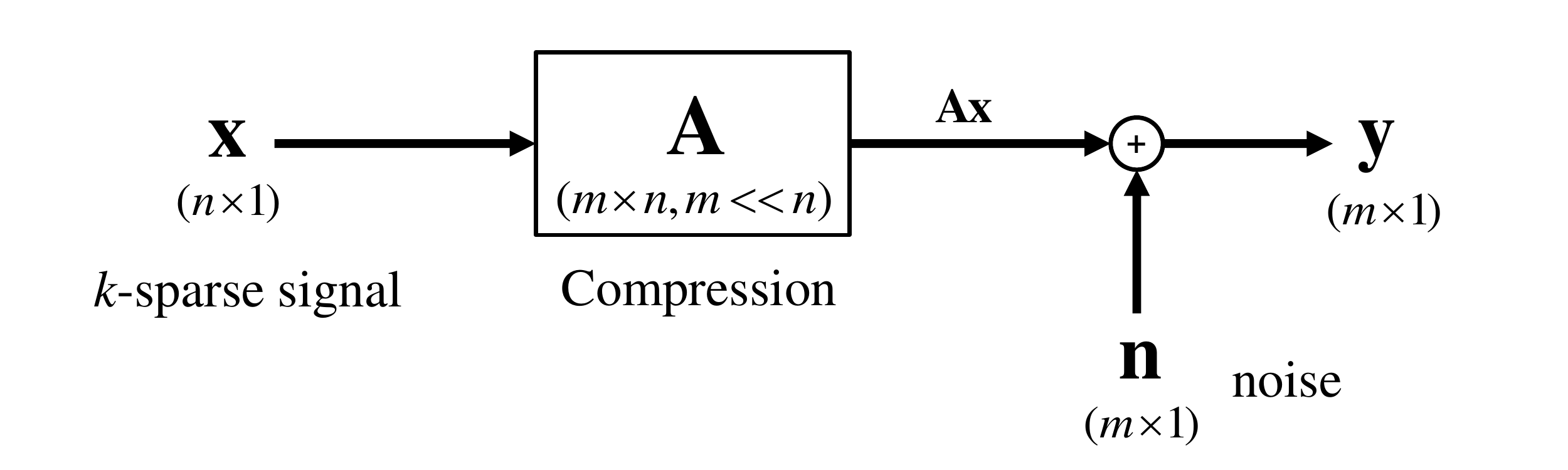}
\caption{Block diagram for compressive sensing process.}
\label{fig:sysmodel}
\end{figure}

\emph{Notation}: We use $\bF_q$ to represent the finite field with $q$ elements, where $q$ is a prime number or power of a prime. For any field $\bF$, we use $\bF[x]$ to denote the polynomial ring in variable $x$ with coefficients from $\bF$. For $n\in\mathbb{N}$, $\vx\in \bF^n$, we use $\textrm{wt}(\vx)$ to denote the number of non-zero elements in $\vx$ (here zero refers to zero element in $\bF$). For $x\in\mathbb{R}$, we use $\lfloor x \rfloor$ and $\lceil x \rceil$ to represent its floor and ceiling values.

Given $b,n,q\in\mathbb{N}$ with $b<n$,  and a finite alphabet $\cA\subset \mathbb{R}$ with $0\in\cA$ and $|\cA|=q$, we consider the following ensemble:
\[
\cS=\{\vx=(x_1,x_2,\ldots,x_n)\in\cA^n:\,\textrm{wt}(\vx)\leq b\}.
\]
This ensemble represents the space of $n$-dimensional signals that are at most $b$-sparse with entries coming from $\cA$. We assume that $q$ is a prime number or its power, and consider a bijective mapping $\phi:\cA\rightarrow \bF_q$ with the restriction $\phi(0)=0$, i.e., $0\in\mathbb{R}$ gets mapped as the zero of $\bF_q$. This allows us to interpret $\cA$ as $\bF_q$ and we define the following set of vectors:
\[
\cS_q=\{\vx=(\phi(z_1),\ldots,\phi(z_n))\in\bF_q^n:\,(z_1,\ldots,z_n)\in\cS\}.
\]
By construction, the vectors in $\cS_q$ are at most $b$-sparse.

We wish to develop a framework for efficient compression of any $\vx\in\mathcal{S}_q$. In precise terms, we desire to reconstruct $\vx\in\cS_q$ using minimal number of measurements that are linear combinations of its elements, based on field operations and coefficients from $\bF_q$. This measurement process is given by
\begin{equation}\label{eqn:sysmodel}
\vy=\mA\vx+\vn,
\end{equation}
where $\mA\in\bF_q^{m\times n}$ is the sensing matrix, $\vn\in\bF_q^m$ is the measurement noise and $\vy\in\bF_q^m$ is the measurement vector. The process is depicted as a block diagram in Figure \ref{fig:sysmodel}. Note that $\vy$ can be interpreted as a noisy and compressed version of $\vx$. The overall goal of the problem setting is to design $\mA$ such that $\vx$ can be recovered accurately and efficiently.

Given the vectors in $\mathcal{S}_q$ are chosen according to an uniform distribution, the application of source coding theorem \cite{ThomCover} states that the number of measurements required to characterize $\vx\in\cS_q$ is at least $\log_2|\mathcal{S}_q|=\log_2[\sum_{j=0}^b{n \choose j}(q-1)^j]=\Omega(b\log(n/b))$ for $b<n/2$. In this paper, we provide schemes for designing $m=2b\lceil \log_q n\rceil$ measurements that permit exact and efficient recovery of $\vx$ from $\vy$. This matches the lower bound on the number of measurements, stated above, in order-wise sense, for the scaling case $b=O(n^\alpha)$, $\alpha\in[0,1)$.

A critical algebraic tool that we utilize for designing the sensing matrices is {\em field lifting}, where we transform the problem from the original finite field to a suitable field extension. We anticipate the audience for this paper may not necessarily be familiar with the algebraic concepts used in this paper. Therefore, to enhance its readability, we set the stage by introducing the concepts of extension fields and field lifting based on extension fields in the subsequent sections.

\subsection{Background: Extension Field}\label{subsec:fieldextension}
We start with the definitions of extension field and subfield:
\begin{defn}[Extension Field]
Let $\bF$ and $\bK$ be fields such that $\bF\subset \bK$. Then $\bK$ is called an extension field of $\bF$ (also denoted by $\bK/\bF$) and $\bF$ is called a subfield of $\bK$.
\end{defn}
As an example, the field of complex numbers $\bC$ is an extension of the field of real numbers $\bR$, constructed using root $i$ of the irreducible polynomial $x^2 + 1$ over $\bR$, as $\bC=\{x+iy:\,x,y\in\bR\}$. In general, irreducible polynomials and their roots play a pivotal role in the generation of extension fields from their base fields; therefore, we define them next:
\begin{defn}[Irreducible Polynomial]
Given a field $\bF$, a polynomial $p(x)\in\bF[x]$ that is divisible only by $cp(x)$ or $c$, $c\in\bF$, is called an irreducible polynomial over $\bF$. Also, if the coefficient of the highest degree term is equal to $1\in\bF$, the polynomial is called monic. A monic irreducible polynomial
with non-trivial degree is called a prime polynomial.
\end{defn}
The extension field $\bF_{q^s}$, $s\in\mathbb{N}$, can be constructed from $\bF_q$ using the root of any irreducible polynomial of degree $s$ over $\bF_q$. It is known that the set of nonzero elements in $\bF_{q^s}$ form a cyclic set, i.e., there exists at least one $\alpha\in\bF_{q^s}$ (called a primitive element of $\bF_{q^s}$) such that $\bF_{q^s}=\{0,1,\alpha,\alpha^2,\ldots,\alpha^{q^s-2}\}$ and $\alpha^{q^s-1}=1$. The existence of primitive elements motivates the concept of primitive polynomial that is defined as follows:
\begin{defn}[Primitive Polynomial]
Given a field $\bF$, a primitive polynomial over $\bF$ (and in $\bF[x]$) is a prime polynomial over $\bF$ having a primitive element of an extension field, say $\bK$, as one of its roots, as a polynomial in $\bK[x]$.
\end{defn}
Using a primitive polynomial $p(x)$ of degree $s$ over $\bF_q$ allows its root to become a generator for the extension field $\bF_{q^s}$, $s\in\mathbb{N}$. Therefore, $\bF_{q^s}$ can be viewed as the set of polynomials over $\bF_q$ modulo $p(x)$. Then $\bF_{q^s}$ can also be viewed as a vector space of dimension $s$ over $\bF_q$, with the basis set $\{1,\alpha,\alpha^2,\ldots,\alpha^{s-1}\}$, where $\alpha$ is the primitive element of $\bF_{q^s}$ such that $p(\alpha)=0$ in $\bF_{q^s}$. As an example, the primitive polynomial $p(x)=x^3+x+1$ over $\bF_2$ and its primitive root $\alpha$ in $\bF_{8}$ can be used to generate the elements of $\bF_{8}$. This field extension process is depicted in Table \ref{table:basis}. The existence of primitive polynomials is confirmed by the following lemma:

\begin{table}
\caption{$\bF_8$ as an extension field of $\bF_2$}
\label{table:basis}
\centering
\begin{tabular}{ccc}
\hline\noalign{\smallskip}
Exponential & Polynomial & Basis $(1\,\,\alpha\,\,\alpha^2)$  \\
\noalign{\smallskip}\hline\noalign{\smallskip}
--         & $0$                 & (0\,\,0\,\,0) \\
$\alpha^0$ & $1$                 & (1\,\,0\,\,0) \\
$\alpha^1$ & $\alpha$            & (0\,\,1\,\,0) \\
$\alpha^2$ & $\alpha^2$          & (0\,\,0\,\,1) \\
$\alpha^3$ & $1+\alpha$          & (1\,\,1\,\,0) \\
$\alpha^4$ & $\alpha+\alpha^2$   & (0\,\,1\,\,1) \\
$\alpha^5$ & $1+\alpha+\alpha^2$ & (1\,\,1\,\,1) \\
$\alpha^6$ & $1+\alpha^2$        & (1\,\,0\,\,1) \\
\noalign{\smallskip}\hline
\end{tabular}
\end{table}

\begin{lem}\label{lem:existence}
Given a finite field $\bF_q$ and $s\in\mathbb{N}$, there are $\varphi(q^s-1)/s$ primitive polynomials over $\bF_q$ that generate the extension field $\bF_{q^s}$. $\varphi(\cdot)$ is referred to as the Euler totient function -- for $n\in\mathbb{N}$ it is defined as $\varphi(n)=n\prod_{i=1}^k\left(1-p_i^{-1}\right)$, where $p_1,p_2,\ldots,p_k$ are prime numbers that divide $n$.
\end{lem}

\subsection{Background: Field Lifting}\label{subsec:fieldlifting}
The idea of field lifting allows one to re-interpret a compressive sensing problem over a particular finite field in one of its extension fields. The motivation for doing so is that extension fields have larger number of dimensions and offer more degrees of flexibility compared to their base fields. Referring to the system model in (\ref{eqn:sysmodel}), field lifting is performed as follows. Given $s\in\mathbb{N}$, we consider a primitive polynomial $p(x)$ of degree $s$ over $\bF_q$ and its root $\alpha$, that is a primitive element of $\bF_{q^s}$. Note that $\bF_{q^s}$ can be viewed as a vector space of dimension $s$ over $\bF_q$  on the basis set $\{1,\alpha,\alpha^2,\ldots,\alpha^{s-1}\}$:
\[
\bF_{q^s}=\left\{\sum_{i=0}^{s-1}c_i\alpha^i:c_0,c_1,\ldots,c_{s-1}\in\bF_q\right\}.
\]
We assume that the number of measurements $m$ satisfies $m=m's$, $m'\in\mathbb{N}$. This allows us to define the following mappings:
\begin{itemize}
\item $\phi_s$: Given $\mC=[\,c_{ij}\,]\in\bF_q^{m\times n}$, $\phi_s(\mC)=[\,c'_{kl}\,]\in\bF_{q^s}^{m'\times n}$ is defined as
    $c'_{kl}=\sum_{t=0}^{s-1}c_{(k-1)s+t+1,l}\alpha^t.$ Thus, the $k$th row of $\phi_s(\mC)$ is obtained by scaling the $((k-1)s+t+1)$th row of $\mC$ by $\alpha^t$, $0\leq t<s$, and summing them up.
\item $\psi_s$: Given $\vc=[\,c_i\,]\in\bF_q^{m}$,  $\psi_s(\vc)=[\,c'_k\,]\in\bF_{q^s}^{m'}$ is defined as $c'_k=\sum_{t=0}^{s-1}c_{(k-1)s+t+1}$$\alpha^t$. Thus, the $k$th entry of $\psi_s(\vc)$ is obtained by scaling the $((k-1)s+t+1)$th entry of $\vc$ by $\alpha^t$, $0\leq t<s$, and summing them up.
\end{itemize}
Note that fixing $p(x)$ and $\alpha$ imparts a unique algebraic structure to $\bF_{q^s}$ in terms of the elements of $\bF_q$ and $\alpha$. As such the mappings $\phi_s$ and $\psi_s$ are bijective functions, i.e., their inverse mappings $\phi_s^{-1}:\bF_{q^s}^{m'\times n}\rightarrow \bF_q^{m\times n}$ and $\psi_s^{-1}:\bF_{q^s}^{m'}\rightarrow \bF_q^{m}$ exist and are well-defined. Assuming measurement noise to be absent in the system (i.e., $\vn=\mathbf{0}$), the system model in (\ref{eqn:sysmodel}) can be restated in terms of $\phi_s$ and $\psi_s$, over $\bF_{q^s}$, as follows:
\begin{equation}\label{eqn:fieldlifting}
\psi_s(\vy)=\phi_s(\mA)\vx.
\end{equation}
Note that $\vx\in\bF_q^n\subseteq\bF_{q^s}^n$. This demonstrates field lifting of the system model from $\bF_q$ to $\bF_{q^s}$. We make use of this concept for designing sensing matrices for compressive sensing over $\bF_q$ in the subsequent sections, using mappings $\phi_s$ and $\psi_s$ for some $s\in\mathbb{N}$, obtained by fixing a primitive polynomial of degree $s$ over $\bF_q$ and choosing a primitive root in $\bF_{q^s}$.

\section{Noiseless Measurements}\label{sec:noiseless}
In this section, we analyze the problem of recovering $\vx\in\cS_q$ in the absence of noise, i.e., $\vn=\mathbf{0}$. Then we have the relation
\begin{equation}\label{eqn:sysmodelnoiseless}
\vy=\mA\vx.
\end{equation}
Note that this situation resembles the process of \emph{syndrome decoding} in linear codes from coding theory, where $\vx,\vy$ and $\mA$ play the roles of error vector, syndrome vector and parity check matrix of the linear code respectively \cite{Roth}. It is this connection to linear codes that we exploit for designing $\mA$ and algorithms for recovering $\vx$ from $\vy$. We refer to a linear code $\mathcal{C}$ as an $[N,K,D]_q$ code ($N,K,D\in\mathbb{N}$ and $q$ is a prime number or its power) if the code alphabet is $\bF_q$, codeword length is $N$, number of codewords is $q^K$, and the minimum Hamming distance between codewords is $D$ (i.e., at most $\lfloor (D-1)/2\rfloor$ errors can be corrected). Then the following theorem holds:
\begin{thm}\label{thm:noiseless}
Given $m=m's$, $m',s\in\mathbb{N}$, it is possible to exactly recover $\vx\in\mathcal{S}_q$ from $\vy$ if $\phi_s(\mA)$ is the parity check matrix of a $[n,n-m',d]_{q^s}$ linear code with $d>2b$.
\end{thm}
\begin{IEEEproof}
We apply mapping $\psi_s$ to obtain $\psi_s(\vy)$. As described in Section \ref{subsec:fieldlifting}, this performs indirect field lifting of the setup from $\bF_q$ to $\bF_{q^s}$, and relation (\ref{eqn:fieldlifting}) holds. Since $\phi_s(\mA)$ is the parity check matrix of a $[n,n-m',d]_{q^s}$ linear code, $\psi_s(\vy)$ acts as a syndrome vector and $\vx$ acts as the error vector generating it. Therefore, syndrome decoding can be used to exactly recover $\vx$ from $\vy$, since the vectors in $\cS_q$ are equally likely to occur (uniform distribution over $\cS_q$), $\textrm{wt}(\vx)\leq b$ and the code can correct up to $\lfloor (d-1)/2\rfloor \geq b$ errors.
\end{IEEEproof}

There exist structured linear codes with efficient algorithms for syndrome decoding -- one example is the family of Reed-Solomon codes \cite{Roth}. Designing $\phi_s(\mA)$ as the parity check matrix of a Reed-Solomon code gives the following corollary:
\begin{cor}\label{cor:noiseless}
Given $m=2bs$, $s\in\mathbb{N}$, $s\geq \lceil \log_q n \rceil$, it is possible to exactly recover $\vx\in\mathcal{S}_q$ from $\vy$ using $O(nbs^2)$ operations in $\bF_q$ if $n>2b$ and $\phi_s(\mA)$ is the parity check matrix of a $[n,n-2b,2b+1]_{q^s}$ Reed-Solomon code.
\end{cor}
\begin{IEEEproof}
The recovery of $\vx$ from $\vy$ follows from Theorem \ref{thm:noiseless} with $m'=2b$. Note that we require $s\geq \lceil \log_q n \rceil$, since the alphabet size should not be less than the codeword length for a Reed-Solomon code, i.e., $q^s\geq n$. Also, there exist multiple algorithms for syndrome decoding in Reed-Solomon codes, such as Euclid's algorithm based decoding or Berlekamp-Massey algorithm, that can reconstruct $\vx$ from $\vy$ using $O(nb)$ operations in $\bF_{q^s}$ \cite{Roth}. Since $\bF_{q^s}$ is a vector space over $\bF_q$, multiplication (and addition) of elements in $\bF_{q^s}$ means multiplication (and addition) of polynomials of degree less than $s$ over $\bF_q$ modulo a primitive polynomial. This implies that a field operation in $\bF_{q^s}$ is equivalent to $O(s^2)$ field operations in $\bF_q$. Overall, this amounts to $O(nbs^2)$ operations in $\bF_q$ for any of the syndrome decoding algorithms.
\end{IEEEproof}
Note that Reed-Solomon codes is one family of codes that can be used for designing sensing matrices. In general, any family of linear codes that admits a polynomial time syndrome decoding algorithm can be used for constructing sensing matrices. Examples include BCH codes and special classes of LDPC codes, like expanders. Also, \eqref{eqn:sysmodelnoiseless} can be thought of as a source coding problem where a vector needs to be compressed. LDGM codes is a family of good source codes that enable efficient vector compression \cite{Frias}. The generator matrix of these codes have the interesting property that vectors close to each other in Hamming distance get mapped to compressed versions that are close in terms of Hamming distance as well. This is, in some sense, analogous to the property of RIP, where low-dimensional projections of vectors, close to each other in $\ell_2$-norm sense, are also mapped close to each other.

{\bf Number of measurements:} Corollary \ref{cor:noiseless} suggests that one can design $m=2b\lceil \log_q n\rceil$ measurements (by choosing $s=\lceil \log_q n \rceil$) for recovering $\vx\in\cS_q$ in the noiseless setting, using $O(nb(\log n)^2)$ field operations in $\bF_q$. This scaling of $m$ is order-wise optimal for $b=O(n^\alpha)$, $\alpha\in[0,1)$, with respect to the information-theoretic lower bound of $\Omega(b\log(n/b))$. A sufficient condition that ensures $\ell_1$-minimization gives accurate recovery in the real-valued framework is that the sensing matrix satisfies RIP of order $2b$ with parameter $\delta_{2b}<(\sqrt{2}-1)$ \cite{CandesTao05}. A convenient way of generating RIP matrices is by choosing its entries from a sub-Gaussian distribution in an i.i.d. fashion. Given $\delta\in(\sqrt{2}-1,1)$ and any $\kappa_1>0$, if $m\geq 2\kappa_1b\log_e(n/2b)$, then exact recovery is possible using $\ell_1$-minimization with probability $\geq 1-2\exp(-\kappa_2 m)$, where $\kappa_2=\delta^2/2\kappa^* - \log_e(42e/\delta)/\kappa_1$ and $\kappa^*=2(1-\log_e 2)$. Therefore, for $b\leq 0.5q^{-(\kappa_1\log_e q)^{-1}}n^{1-(\kappa_1\log_e q)^{-1}}$ (i.e., sparsity below some threshold) the number of measurements required for the finite field/alphabet framework is smaller.

{\bf Storage space:} The storage space needed for the measurement vector is at most $2b \lceil \log_q n \rceil\log_2 q$ bits, that lies between $2b\log_2 n$ and $2b\log_2 n+2b\log_2q$ bits. The storage space taken by real-values is in theory, infinite, and in practice, with $j$-bit quantization, is linear in $j$. For the case of real-valued compressing and the same number of measurements, this amounts to $2jb\lceil\log_q n\rceil$ bits of storage space for the measurement vector. Note that $j> \log_2 q$, since at least $\log_2q$ bits are needed to resolve among the elements of finite alphabet of size $q$ (even if they are treated as real numbers). This gives storage space of at least $2b \lceil \log_q n \rceil\log_2 q$ bits. Therefore, the storage requirement for the finite field framework is smaller compared to the real-valued framework -- also, the ratio of the number of bits needed for the algebraic approach vs. the real-valued approach is $\approx \log_2 q/j$.

Thus, the algebraic approach offers benefits in terms of lesser number of measurements and storage space (in bits), provided the sparsity levels are below some threshold.

\section{Noisy Measurements}\label{sec:noisy}
In this section, we analyze the problem of recovering $\vx\in\mathcal{S}_q$ in presence of noise; the purpose being to demonstrate the versatility and utility of coding-theoretic tools for finite alphabet sparse signal recovery. We consider two noise models for analysis -- probabilistic noise and worst-case noise.

\subsection{Probabilistic Noise Model}
The probabilistic noise model is widely used for modeling errors resulting from transmissions across communication channels. For this model, we assume that $\vn$ is generated according to a probability distribution. For the sake of simplicity, we consider $\vn$ being generated by $m$ independent uses of a $q$-ary symmetric channel with crossover probability $\lambda\in(0,1-q^{-1})$ (similar analysis can be done for noise with general probability distributions). In other words, if $\vn=[n_1\,\,n_2\,\cdots\, n_m]^T$, $n_i$ has the probability distribution $P(n_i=a)=\lambda/(q-1)$ for $a\in\bF_q\backslash\{0\}$ and $1-\lambda$ for $a=0$. Here, we recover $\vx$ from $\vy$ in two steps. First, we eliminate the effect of errors introduced by $\vn$, using error correction capability of linear codes, and obtain a compressed version of $\vx$, with high probability. Next, we retrieve $\vx$ from this compressed version, as described in Section \ref{sec:noiseless}.

We say that a linear code $\mathcal{C}$ achieves probability of error of at most $P_e$ over a channel if $\max_{\mathbf{c}\in\mathcal{C}}P_e(\mathbf{c})\leq P_e$, where $P_e(\mathbf{c})$ refers to the probability that codeword $\mathbf{c}$ is decoded erroneously by a nearest-neighbor-codeword decoder, conditioned on the fact that $\mathbf{c}$ was originally sent across the channel. We also define $H_q(x)\triangleq-x\log_qx-(1-x)\log_q(1-x)+x\log_q(q-1)$, $x\in(0,1)$. Then the following theorem holds:
\begin{thm}\label{thm:noisy1}
Given $\lambda\in(0,1-q^{-1})$, $m=cm'\geq cm''s$, $c,m',m'',s\in\mathbb{N}$, $c>1/(1-H_q(\lambda))$, it is possible to exactly recover $\vx\in\mathcal{S}_q$ from $\vy$ with probability $\geq (1-P_e)$ if $\mA=\mG\mA'$, where $\mG$ is the generator matrix of a $[m,m',d]_q$ linear code, achieving probability of error of at most $P_e$ over $q$-ary symmetric channel with crossover probability $\lambda$, and some set of $m''s$ rows of $\mA'$ forms $\mA''$  such that $\phi_s(\mA'')$ is the parity check matrix of a $[n,n-m'',d']_{q^s}$ linear code, $d'>2b$.
\end{thm}
\begin{IEEEproof}
Using the given form of $\mA$, we have $\vy=\mG\mA'\vx+\vn.$ Since $\mG$ is the generator matrix of a $[m,m's,d]_q$ linear code, $\mG\mA'\vx$ can be treated as a codeword, $\vy$ as its noisy version and $\mA'\vx$ as the message vector generating the codeword. Therefore, it is possible to recover $\mA'\vx$ from $\vy$ with probability $\geq (1-P_e)$. Note that the restriction on $c$ arises from the fact that the rate of the linear code corresponding to $\mG$, i.e., $1/c$, cannot exceed $1-H_q(\lambda)$, the capacity of the $q$-ary symmetric channel. The knowledge of $\mA'\vx$ gives $\mA''\vx$ since the rows of $\mA''$ form a subset of the rows of $\mA'$. Thereafter, $\vx$ can be recovered from $\mA''\vx$ as described in the proof of Theorem \ref{thm:noiseless} (via field lifting and syndrome decoding).
\end{IEEEproof}

Note that $\phi(\mA'')$ can be chosen as the parity check matrix of a Reed-Solomon code and $\mA'$ can be designed to have $\mA''$ as its sub-matrix. One family of linear codes that achieves small error probabilities over $q$-ary symmetric channel is concatenated codes \cite{Roth,Dumer}. For example, given $\lambda\in(0,1-q^{-1}),\epsilon\in(0,1-H_q(\lambda)),\rho\in(0,1)$ and large enough $t\in\mathbb{N}$, one can design a concatenated $[N,K,D]_q$ linear code with $N=tq^{\lfloor\epsilon t\rfloor}$ and $K=\lfloor \epsilon t\rfloor \lceil\rho q^{\lfloor\epsilon t\rfloor}\rceil$, whose decoding algorithm requires $O(N^2\log N)$ operations in $\bF_q,\bR$ \cite{Roth}. Furthermore, the code achieves probability of error of at most $q^{-c(\epsilon,\rho)N}$ over a $q$-ary symmetric channel with crossover probability $\lambda$, where $c(\epsilon,\rho)$ is a positive constant dependent only on $q,\epsilon,\rho$. We refer to this linear code as $\mathcal{C}_{con}(t,\epsilon,\rho)$. Designing $\mG$ as the generator matrix of this code gives the following corollary:
\begin{cor}
Given $\lambda\in(0,1-q^{-1})$, $\rho\in(0,1)$, $\epsilon\in(0,1-H_q(\lambda))$, $s\in\mathbb{N}$, $s\geq \lceil \log_q n \rceil$, it is possible to exactly recover $\vx\in\mathcal{S}_q$ from $\vy$ with probability $\geq (1-q^{-cbs})$ ($c>0$ is dependent on $q,\epsilon,\rho$) for sufficiently large $b,n$ using $O((n+b\log(bs))bs^2)$ operations in $\bF_q$ and $\bR$ if $n>2b$ and $\mA=\mG\mA'$, where $\mG$ is the generator matrix of $\mathcal{C}_{con}(\lceil t^*\rceil,\epsilon,\rho)$, $t^*\in\bR$ being the solution to  $\epsilon\rho xq^{\epsilon x}=4qbs$, and some set of $2bs$ rows of $\mA'$ forms $\mA''$  such that $\phi_s(\mA'')$ is the parity check matrix of a $[n,n-2b,2b+1]_{q^s}$ Reed-Solomon code.
\end{cor}
\begin{IEEEproof}
The recovery of $\vx$ from $\vy$ with probability $\geq (1-q^{-c bs})$, for some constant $c>0$, follows from Theorem \ref{thm:noisy1} with $m''=2b$, the properties of codebook $\mathcal{C}_{con}(\lceil t^*\rceil,\epsilon,\rho)$ and the fact that the number of rows of $\mA'$ is $\lfloor\epsilon\lceil t^*\rceil\rfloor\lceil q^{\lfloor \epsilon \lceil t^*\rceil \rfloor}\rceil$, that is bounded below by $2bs$ and bounded above by $16q^{1+\epsilon}bs$, for large enough values of $b,n$. By property of the code $\mathcal{C}_{con}(\lceil t^*\rceil,\epsilon,\rho)$, $\mA'\vx$ is recoverable from $\vy$ using $O((\lceil t^*\rceil q^{\epsilon \lceil t^*\rceil})^2\log(\lceil t^*\rceil q^{\epsilon \lceil t^*\rceil}))=O(b^2s^2\log(bs))$ operations in $\bF_q$ and $\bR$. The knowledge of $\mA'\vx$ gives $\mA''\vx$ since the rows of $\mA''$ form a subset of the rows of $\mA'$. Thereafter, $\vx$ can be obtained from $\mA''\vx$ via field lifting and syndrome decoding using $O(nbs^2)$ operations in $\bF_q$, as described in the proof of Corollary \ref{cor:noiseless}. This amounts to a total of $O((n+b\log(bs))bs^2)$ operations in $\bF_q,\bR$.
\end{IEEEproof}

The above corollary suggests that one can design $m=\Theta(b\log_q n)$ measurements (by choosing $s=\lceil \log_q n \rceil$) for recovering any $\vx\in\cS_q$ in presence of $q$-ary symmetric noise for large enough values of $b,n$, using $O((n+b\log b+b\log\log n)b(\log n)^2)$ field operations in $\bF_q,\bR$. This scaling of $m$ is order-wise optimal for $b=O(n^\alpha)$, $\alpha\in[0,1)$, with respect to the information-theoretic lower bound. Note that there are no theoretical guarantees in the context of real-valued compressing sensing for noise model; most of the guarantees are designed for Gaussian and $\ell_2$-norm bounded noise.

\subsection{Worst-case Noise Model}
The worst-case noise model has been used for modeling corruption in storage media as well as channel transmission errors. Here, we assume that $\vn$ comes from the following ensemble of signals with bounded number of non-zero entries:
\[
\mathcal{N}(\delta)=\{\vn\in\bF_q^m:\textrm{wt}(\vn)\leq \delta m\},\quad 0<\delta<1/2.
\]
For the sake of simplicity, we assume that $\delta m\in\mathbb{N}$. Here, we recover $\vx$ from $\vy$ in two steps, similar to the recovery procedure for the probabilistic noise model. First, we eliminate the errors introduced by $\vn$ using the error correction capability of linear codes, and obtain a noiseless compressed version of $\vx$. Next, we reconstruct $\vx$ from this version using the approach described in Section \ref{sec:noiseless}. The following theorem holds here:

\begin{thm}\label{thm:noisy2}
Given $\delta\in(0,(1-q^{-1})/2)$, $\vn$ coming from $\mathcal{N}(\delta)$, $m>m'\geq m''s$, $m',m'',s\in\mathbb{N}$, it is possible to exactly recover $\vx\in\mathcal{S}_q$ from $\vy$ if $\mA=\mG\mA'$, where $\mG$ is the generator matrix of a $[m,m',d]_q$ linear code, $d>2\delta m$, and some set of $m''s$ rows of $\mA'$ forms $\mA''$  such that $\phi_s(\mA'')$ is the parity check matrix of a $[n,n-m'',d']_{q^s}$ linear code, $d'>2b$.
\end{thm}
\begin{IEEEproof}
Using the given form of $\mA$, we have $\vy=\mG\mA'\vx+\vn.$ Since $\mG$ is the generator matrix of a $[m,m',d]_q$ linear code, $\mG\mA'\vx$ can be treated as a codeword and $\mA'\vx$ can treated as the message vector generating it. Therefore, a decoding algorithm can be used to recover $\mA'\vx$ from $\vy$, since $\textrm{wt}(\vn)\leq \delta m$ and the linear code can correct up to $\delta m$ errors. Note that the restriction $\delta\in(0,(1-q^{-1})/2)$ arises due to Plotkin's bound for linear codes \cite{Roth}. The knowledge of $\mA'\vx$ gives $\mA''\vx$ since the rows of $\mA''$ form a subset of the rows of $\mA'$. Thereafter, $\vx$ can be recovered, as described in Theorem \ref{thm:noiseless}.
\end{IEEEproof}

There exist families of linear codes with good minimum distance properties, like concatenated codes. In particular, given $\epsilon,\rho\in(0,1)$ and large enough $t\in\mathbb{N}$, it is possible to design a concatenated $[N,K,D]_q$ linear code with $N=tq^{\lfloor\epsilon t\rfloor}$, $K=\lfloor \epsilon t\rfloor \lceil\rho q^{\lfloor\epsilon t\rfloor}\rceil$, $D\geq (H_q^{-1}(1-\epsilon))(1-\rho)tq^{\lfloor\epsilon t\rfloor}$. We refer to this linear code as $\mathcal{D}_{con}(t,\epsilon,\rho)$; also, its decoding process requires $O(N^2\log N)$ operations in $\bF_q$ and $\bR$ \cite{Roth}. Designing $\mG$ as its generator matrix results in the following corollary:
\begin{cor}
Given $\rho\,\,\in\,\,(0,1)$, $\delta\,\,\in\,\,(0,(1-q^{-1})(1-\sqrt{\rho})/2)$, $\epsilon\,\,= \,\,1-H_q(2\delta/(1-\sqrt{\rho}))$, $\vn$ coming from $\mathcal{N}(\delta)$, $s\in\mathbb{N}$, $s\geq \lceil \log_q n \rceil$, it is possible to exactly recover $\vx\in\mathcal{S}_q$ from $\vy$ for sufficiently large $b,n$ using $O((n+b\log(bs))bs^2)$ operations in $\bF_q$ and $\bR$ if $n>2b$ and $\mA=\mG\mA'$, where $\mG$ is chosen as the generator matrix of $\mathcal{D}_{con}(\lceil t^*\rceil,\epsilon,\rho)$, $t^*\in\bR$ being the solution to $\epsilon\rho xq^{\epsilon x}\,=\,4qbs$, and some set of $2bs$ rows of $\mA'$ forms $\mA''$ such that $\phi_s(\mA'')$ is the parity check matrix of a $\,\,[n,n-2b,2b+1]_{q^s}\,\,$ Reed-Solomon code.
\end{cor}
\begin{IEEEproof}
The recovery of $\vx$ from $\vy$ follows from Theorem \ref{thm:noisy2} with $m''=2b$, the properties of codebook $\mathcal{D}_{con}(\lceil t^*\rceil,\epsilon,\rho)$ and the fact that the number of rows of $\mA'$ is $\lfloor\epsilon\lceil t^*\rceil\rfloor\lceil q^{\lfloor \epsilon \lceil t^*\rceil \rfloor}\rceil$, that is bounded below by $2bs$ and bounded above by $16q^{1+\epsilon}bs$, for large values of $b,n$. By property of the codebook, $\mA'\vx$ is recoverable from $\vy$ in  $O((\lceil t^*\rceil q^{\epsilon \lceil t^*\rceil})^2\log(\lceil t^*\rceil q^{\epsilon \lceil t^*\rceil}))=O(b^2s^2\log(bs))$ operations in $\bF_q,\bR$. The knowledge of $\mA'\vx$ gives $\mA''\vx$ since the rows of $\mA''$ form a subset of the rows of $\mA'$. Then $\vx$ can be obtained from $\mA''\vx$ via syndrome decoding using $O(nbs^2)$ operations in $\bF_q$ using Corollary \ref{cor:noiseless}. This amounts to a total of $O((n+b\log(bs))bs^2)$ operations in $\bF_q$ and $\bR$.
\end{IEEEproof}

The above corollary suggests that one can design $m=\Theta(b\log_q n)$ measurements (by choosing $s=\lceil \log_q n \rceil$) for recovering any $\vx\in\cS_q$ in the worst-case noise setting for large enough $b,n$, using $O((n+b\log b+b\log\log n)b(\log n)^2)$ field operations in $\bF_q$, provided the fraction of corrupted symbols $\delta$ is $<(1-q^{-1})/2$. This scaling of $m$ is order-wise optimal for $b=O(n^\alpha)$, $\alpha\in[0,1)$, with respect to the information-theoretic lower bound. Note that similar to probabilistic noise model, there are no theoretical guarantees for real-valued compressing sensing based on worst-case noise.

\section{Simulation Results}\label{sec:simulation}
In this section, we present simulation results showing the utility of our approach in the context of sensing and recovering sparse data, and tracking discrete-valued time series.

\begin{figure*}[t]
\centering
\subfloat[]{\includegraphics[width=3in]{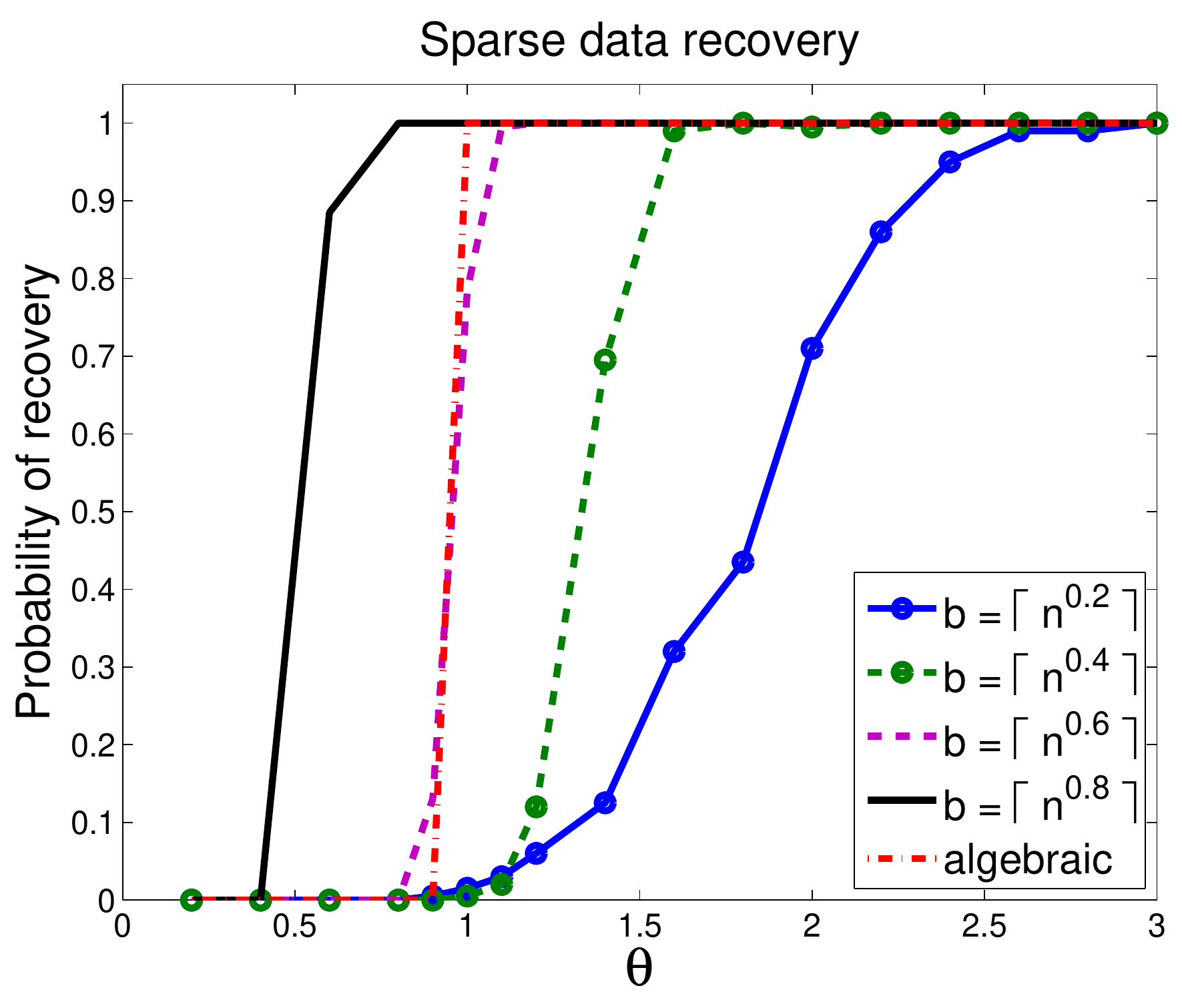}\label{fig:plot_1024}}\hspace{0.75in}
\subfloat[]{\includegraphics[width=3in]{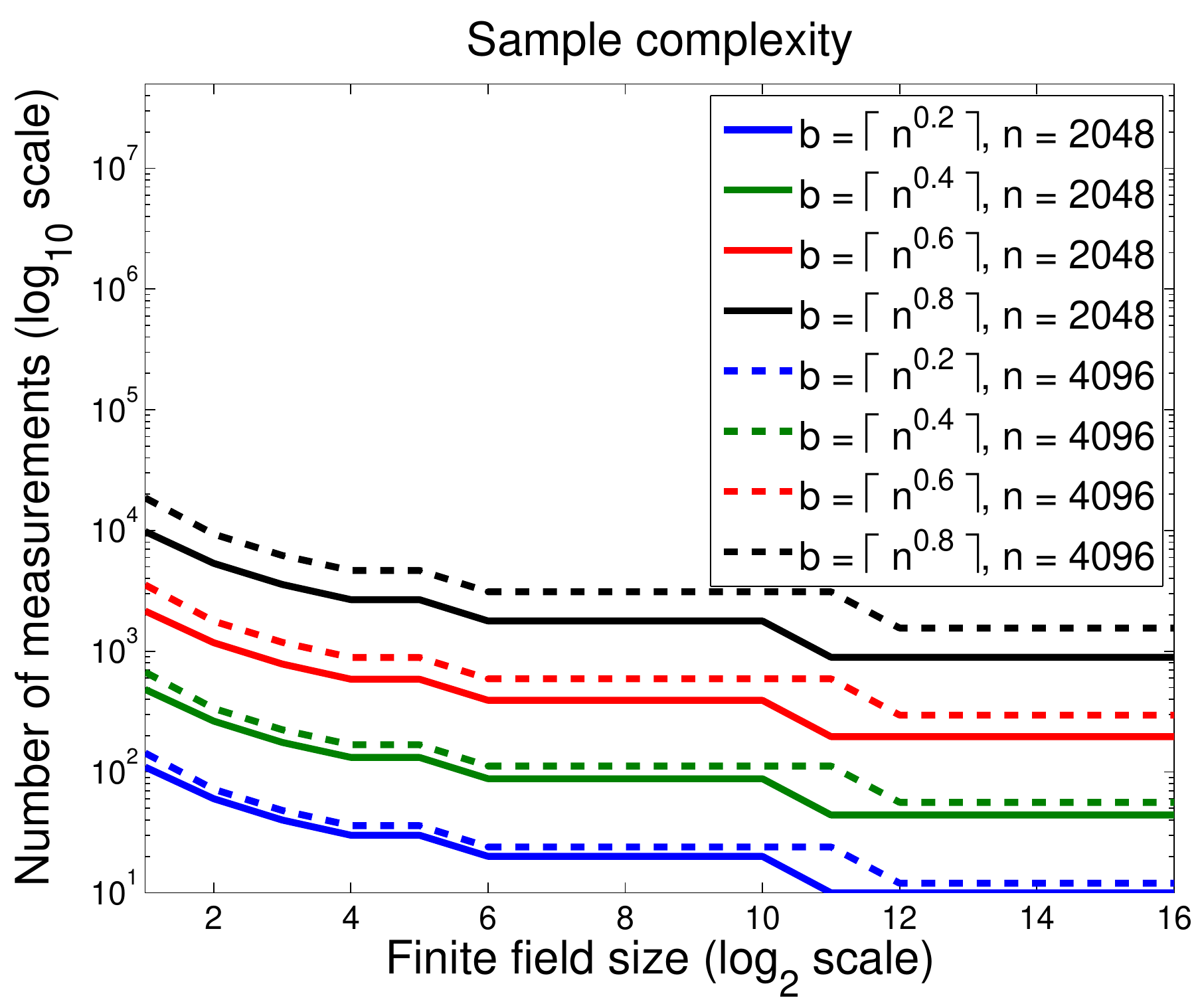}\label{fig:plot_complexity}}\\
\subfloat[]{\includegraphics[width=3in]{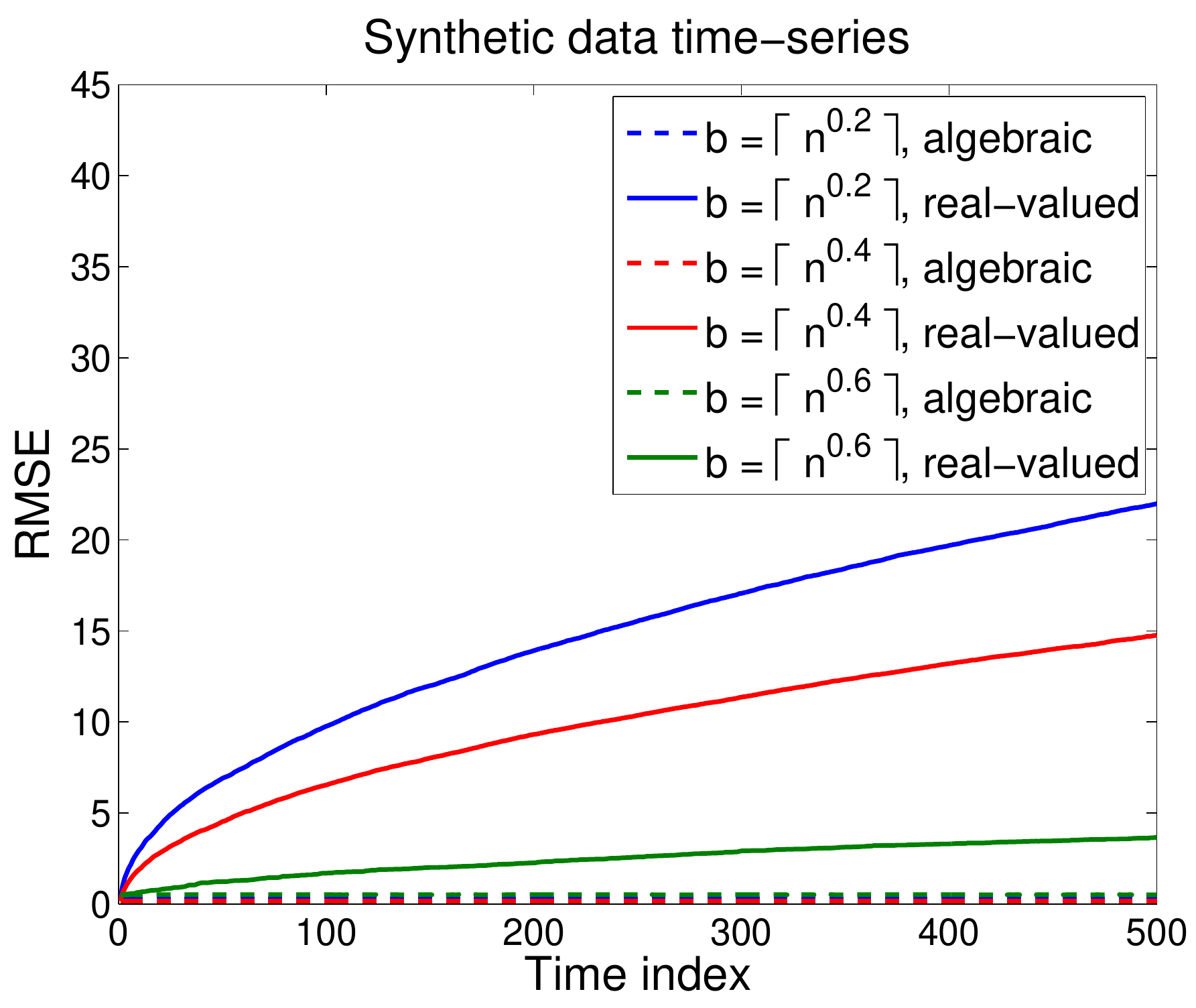}\label{fig:data_tracking}}\hspace{0.75in}
\subfloat[]{\includegraphics[width=3in]{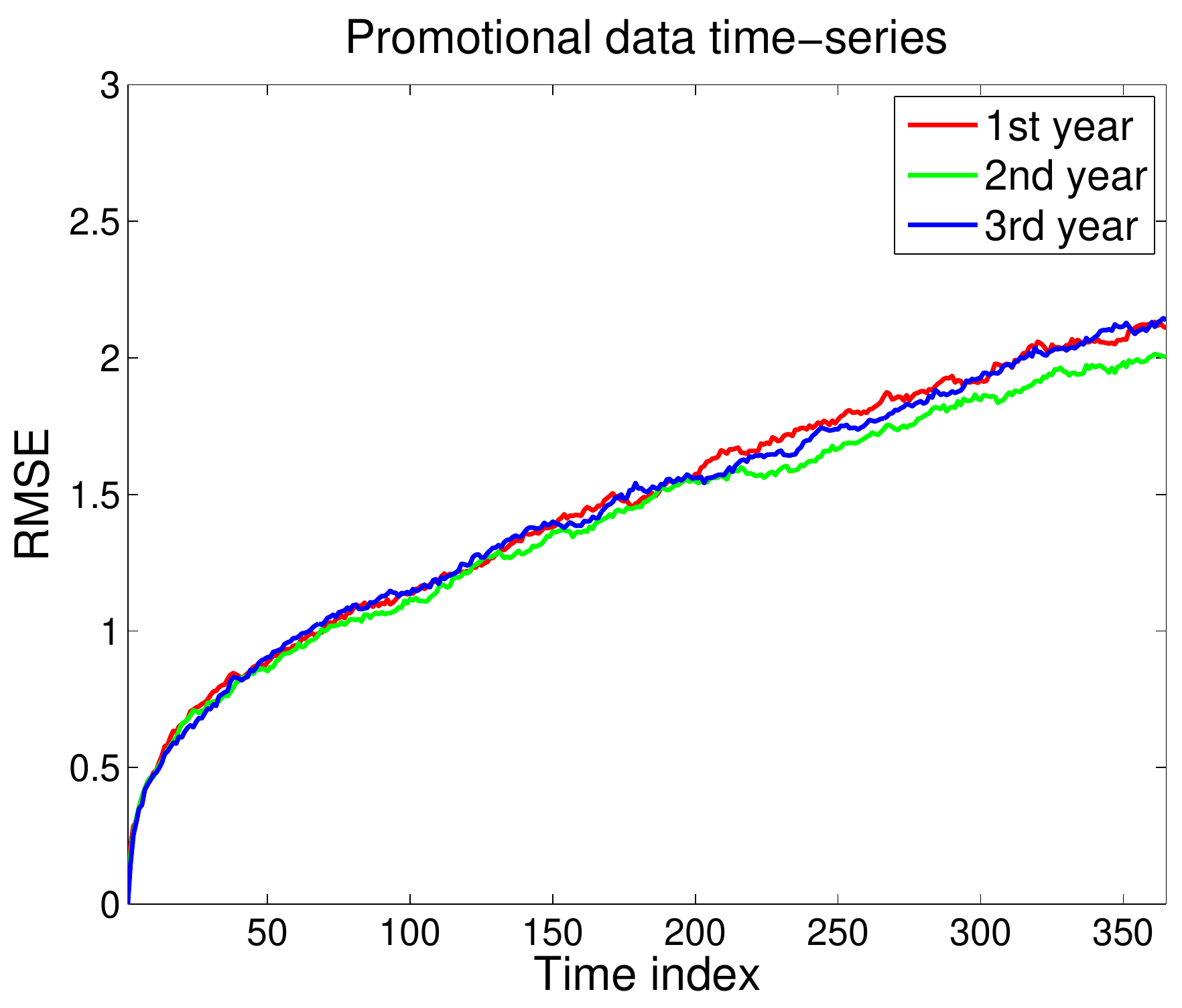}\label{fig:promo_tracking}}\\
\caption{Simulation results for (a),(b) synthetic sparse data, and (c),(d) discrete-valued time series.}
\end{figure*}

{\bf Synthetic sparse data:} We show the effect of different sparsity levels on our approach vs. real-valued compressive sensing using synthetically generated sparse data. We choose $n=1024$ and four sparsity levels, $b=\lceil n^r \rceil$, $r=0.2$, $0.4$, $0.6$, $0.8$. We set the number of linear measurements as $m=2\lfloor\theta b\rfloor \lceil\log_q n\rceil$, where $0.2\leq \theta\leq 3$ and $q=256$.

For real-valued compressive sensing, we take $200$ realizations of $b$-sparse vector $\vx\in\bR^n$, by choosing $b$ positions of a $n$-length zero vector uniformly at random and setting the entries to $1\in\bR$. We construct $\mA\in\bR^{m\times n}$ by choosing its entries in an i.i.d. fashion from the Gaussian distribution with mean $0$ and variance $1/\sqrt{m}$. We reconstruct $\hat{\vx}$, the estimate for $\vx$, using $\ell_1$-minimization and define the error-free event as $\{||\vx-\hat{\vx}||_2/\sqrt{n}<10^{-3}\}$. The probability of recovery is defined as the number of times the error-free event occurs divided by the number of sparse realizations, i.e, 200.

For the algebraic approach, we take the same sparse vector realizations, but treat their entries as elements of $\bF_{q}$, ($0$ is treated as the zero element of the field and $1$ is treated as the identity element of $\bF_q$). We compress and recover the sparse vectors using the approach described in Section \ref{sec:noiseless}. With $s=\lceil\log_{q} n\rceil$, we choose sensing matrix $\mA\in\bF_{q}^{m\times n}$ such that $\phi_s(\mA)$ is the parity check matrix of a $[n,n-2\lfloor\theta b\rfloor, 2\lfloor\theta b\rfloor+1]_{q^s}$ Reed-Solomon code; this ensures that the number of measurements is $m=2\lfloor\theta b\rfloor \lceil\log_q n\rceil$. We also define the error-free event as the case of exact recovery of $\vx$ and the probability of recovery as the number of times exact recovery occurs divided by the number of sparse realizations chosen, i.e, 200.

Figure \ref{fig:plot_1024} shows the plot of $\theta$ vs. probability of recovery. The fact that the number of measurements required for the algebraic approach is lesser compared to that of real-valued compressive sensing for sparsity levels $b=\lceil n^r \rceil$, $r=0.2$, $0.4$, $0.6$, corroborates the remark made about sample complexity in Section \ref{sec:noiseless}. This also implies lesser storage space for measurements and sensing matrices for cases of low sparsity levels, since an element in $\bF_q$ can be represented in $\log_2 q=8$ bits whereas reals are generally assigned more bits (assigning lesser bits would give quantization error as overhead).

Note that the sufficient condition of $m=2b\lceil\log_q n\rceil$ measurements for the algebraic approach implies the required number of measurements decreases as the field size increases. We demonstrate this fact in Figure \ref{fig:plot_complexity}, where we consider compressive sensing setup over finite alphabet represented by $\bF_{q}$, $q=2^i$, $i=1,2,\ldots,16$, with $n=2048$, $4096$ and $b=\lceil n^r \rceil$, $r=0.2$, $0.4$, $0.6$, $0.8$. This figure shows the plot of $\log_2 q$ (bit-resolution of $\bF_q$) vs. $m=2b\lceil\log_q n\rceil$. One can observe that the number of measurements saturates to $2b$ for large values of $q$, a lower bound on the sample complexity for differentiating between two $b$-sparse vectors or signals.

%\begin{figure*}[t]
%\centering
%\subfloat[]{\includegraphics[width=2.5in]{Fig3.pdf}\label{fig:data_tracking}}\hspace{1in}
%\subfloat[]{\includegraphics[width=2.5in]{Fig4.pdf}\label{fig:promo_tracking}}\\
%\caption{Simulation results for tracking discrete-valued time series.}
%\end{figure*}

{\bf Tracking discrete-valued time series:} The problem of the tracking time series is an important one and has been well-studied in literature \cite{Chen,Concha}. In many situations, the changes in the variable associated with the time series is sparse, such as sequence of video frames and time series from human motion recognition or animation. As such, the concept of compressive sensing can be used -- the idea is to compress the sparse changes in the variable to reduce the amount of memory needed for storing the time-series information.

As an example, consider a real-valued time series that has been quantized to get a discrete-valued time series $(\vz_1,\vz_2,\ldots,\vz_t)$, where $\vz_i\in\cA^n$ and $\cA\subset\bR$ is the discrete alphabet, with the property $\textrm{wt}(\ve_i)\leq b$, $\ve_i\triangleq \vz_{i+1}-\vz_i$, $i=1,2,\ldots,t-1$, and $b<<n$. Then one approach to compress the time series is to use real-valued compressive sensing --  consider a sensing matrix $\mA\in\bR^{m\times n}$, satisfying some incoherence property and compress/track the discrete-valued time series as $(\vz_1,\mA\ve_1,\ldots,\mA\ve_{t-1})$. The decompression algorithm comprises of recovering $\ve_1,\ve_2,\ldots,\ve_{t-1}$ using $\ell_1$-minimization and getting the estimate of the discrete-valued time series. Another approach is the algebraic one -- interpret $\cA$ as finite field or subset of a finite field and perform compression using field operations, using the methods described in Section \ref{sec:noiseless}. Next, we provide simulation results for the tracking error of a synthetically generated quantized time series and a real promotion data based time series.

We consider the parameter values $n=1024$, $t=500$ and sparsity levels $b=\lceil n^r \rceil$, $r=0.2$, $0.4$, $0.6$, for generating the synthetic time series $(\vx_1,\vx_2,\ldots,\vx_t)$, $\vx_i\in\bR^n$, as follows. We construct $\vx_1$ by selecting its real-valued entries uniformly at random from $[-1,1]$. We construct $\vf_i$ from the $n$-length zero vector by selecting $b$ random positions in $\vf_i$ and replacing the entries in those positions with uniformly selected real numbers from $[-1,1]$, $i=1,2,\ldots,t-1$. This gives us the desired time series with $\vx_{i+1}\triangleq \vx_{i}+\vf_i$, $i=1,2,\ldots,t-1$. For quantizing this time-series, we use a simple approach -- we choose the minimum and maximum values, say $m$ and $M$ respectively, an entry in the time series takes, and perform quantization in steps of $\delta=(M-m)/q$, $q=256$. Therefore, the quantized time series, say $(\vz_1,\vz_2,\ldots,\vz_t)$, is based on a finite alphabet of size $q$. We define $\ve_i\triangleq \vz_{i+1}-\vz_i$, $i=1,2,\ldots,t-1$, that are $b$-sparse by construction, and set $m=2b\lceil \log_q n\rceil$.

For the real-valued compressive sensing approach, we perform compression and tracking of this quantized series using a sensing matrix $\mA\in\bR^{m\times n}$, whose entries are i.i.d. entries from the Gaussian distribution with mean $0$ and variance $1/\sqrt{m}$. We recover $\ve_1,\ve_2,\ldots,\ve_{t-1}$ from $\mA\ve_1,\mA\ve_2,\ldots,\mA\ve_{t-1}$ using $\ell_1$-minimization, obtain estimates $(\hat{\ve}_1,\ldots,\hat{\ve}_{t-1})$ and estimate the quantized time series as $(\vz_1,\hat{\vz}_2,\ldots,\hat{\vz}_{t})$, $\hat{\vz}_{i+1}=\hat{\vz}_i+\hat{\ve}_i$, $i=1,2,\ldots,t-1$. We define the tracking error at time $i$ as $||\vx_i-\hat{\vz}_i||_2/\sqrt{n}$. For the algebraic approach, we treat the quantization alphabet as $\bF_q$ and use the sensing matrix construction in Section \ref{sec:noiseless}, based on Reed-Solomon codes. Since $m=2b\lceil \log_q n\rceil$, we have exact recovery in this case, so the tracking error at time $i$ only comprises of the quantization error $||\vx_i-\vz_i||_2/\sqrt{n}$. Figure \ref{fig:data_tracking} shows the plot for tracking error vs. time index for different sparsity levels. Note that this increasing nature of tracking error for real-valued compressive sensing is due to error propagation in the estimates of the time series; this includes both the quantization error as well as the error in determining the sparsity patterns of the changes in the time-series vector variable. Also, the tracking error reduces with increase in $b$, since $m$ increases and approaches closer to the optimal number of measurements required for real-valued compressive sensing for error-free recovery of the sparse variable changes.

The promotional data time series comes from \cite{PROMO}. We make use of the 'promotions.dat' file containing a time series with vector-length as $n=1000$ and number of time indices as $1000$. The time index refers to the day index (so it is 3 years of data) and the vector entries refer to the products. The entries in the time series come from $\{0,1\}$, $1$ means the product was promoted that day and $0$ means no promotion for the product. We first consider data for the first year for our simulations, $t=1$ to $t=365$. We refer to this time series as $(\vz_1,\vz_2,\ldots,\vz_T)\in\{0,1\}^n$, $T=365$. We observe that the Hamming distance between two successive vectors never exceeds $60$. In other words, the changes in the support indices of successive vectors are $b$-sparse, $b=60$. We perform the same manner of tracking/compression as we do for the synthetic time series, the only difference being that the promotional time series is already discrete-valued and hence there is no need for quantization. We set the number of measurements as $m=2b\lceil \log_q n\rceil$, $q=1024$. For the algebraic approach, we treat $\{0,1\}$ as a subset of $\bF_q$ ($0$ is treated as the zero element of the field and $1$ is treated as the identity element of the field) and use the sensing matrix construction in Section \ref{sec:noiseless}. Since $m=2b\lceil \log_q n\rceil$, the tracking error is always zero for the algebraic approach over $\bF_q$, i.e., we have perfect tracking. Figure \ref{fig:promo_tracking} shows the tracking error with increasing time index. Note that the tracking error for compressive sensing over reals increases with time due to error propagation in the estimates of the time series. We repeat the same procedure for time series data for the second and third years. Also, note that the algebraic approach requires lesser amount of storage space if a real number is assigned $\log_2 q=10$ bits or more.

\section{Conclusion}\label{sec:conclusion}
In this paper, we develop an algebraic framework for compressive sensing over finite alphabet; this bridges the areas of coding theory and compressive sensing in a thought-provoking way. In particular, it give us tools and constructive approaches for designing sensing matrices as well as polynomial-time-complexity algorithms for sparse source recovery, all while maintaining optimality in terms of sample complexity for exact recovery. Furthermore, we demonstrate that our approach outperforms real-valued compressive sensing in terms of sample complexity and storage space if the sparsity level is below some threshold, and is extendible to the case of noisy measurements, with respect to a broad range of noise models. In terms of utility, finite alphabet compressive sensing proves to be a natural fit for the purpose of compressing/tracking discrete-valued time-series data with sparse changes.

\bibliographystyle{IEEEbib}
\bibliography{refs}

\end{document}